\documentclass{INTERSPEECH2023}
\interspeechcameraready
\usepackage{subcaption}
\usepackage[table,xcdraw]{xcolor}
\usepackage{hyperref}
\newcommand{\squeezeup}{\vspace{-2.0mm}}

\title{Whisper-AT: Noise-Robust Automatic Speech Recognizers \\are Also Strong General Audio Event Taggers}
\name{
Yuan Gong$^1$, Sameer Khurana$^1$, Leonid Karlinsky$^2$, James Glass$^1$}
\address{
\vspace{-2mm}
  $^1$MIT CSAIL, USA \quad\quad $^2$MIT-IBM Watson AI Lab, USA
\email{\{yuangong,glass\}@mit.edu \\ \vspace{2mm}
\href{https://github.com/YuanGongND/whisper-at}{\color{blue}{\texttt{github.com/yuangongnd/whisper-at}}} }
}
\begin{document}

\maketitle
 
\begin{abstract}
In this paper, we focus on Whisper~\cite{radford2022robust}, a recent automatic speech recognition model trained with a massive 680k hour labeled speech corpus recorded in diverse conditions. We first show an interesting finding that while Whisper is very robust against real-world background sounds (e.g., music), its audio representation is actually not noise-invariant, but is instead highly correlated to non-speech sounds, indicating that Whisper recognizes speech \emph{conditioned} on the noise type. With this finding, we build a unified audio tagging and speech recognition model \emph{Whisper-AT} by freezing the backbone of Whisper, and training a lightweight audio tagging model on top of it. With  $<$1\% extra computational cost, Whisper-AT can recognize audio events, in addition to spoken text, in a single forward pass.
\end{abstract}

\vspace{-0.7mm}
\section{Introduction}
\vspace{-1.0mm}
% asr has broad applications

%Automatic speech recognition (ASR) is an active research area.
In recent years, significant progress has been made in advancing automatic speech recognition (ASR) performance. Specifically, self-supervised learning schemes such as wav2vec2.0~\cite{baevski2020wav2vec} and Hubert~\cite{hsu2021hubert} have achieved great success, requiring minimal \emph{labeled} training data. However, since the public model checkpoints are trained with clean speech data (e.g., Librispeech~\cite{panayotov2015librispeech} or Libri-light~\cite{kahn2020libri}), their robustness in real-world environments is limited. To improve noise robustness, the Whisper~\cite{radford2022robust} model uses 680K hours of \emph{labeled} speech collected from the Internet with \emph{diverse} environments and recording setups as the training data, and reports better robustness over existing ASR models.

%unlike wav2vec2 and Hubert, 
%In consequence, Whisper reports lower word error rates (WER) under white and pop noises and shows stronger performance in the real world. 

\begin{figure}[!t]
\centering
\begin{subfigure}[t]{0.43\textwidth}
\includegraphics[width=\textwidth]{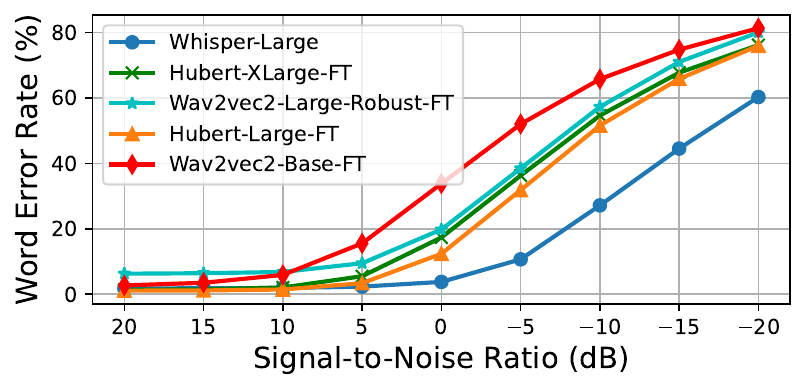}
\end{subfigure}
\vspace{-1.8mm}

\begin{subfigure}[t]{0.43\textwidth}
\includegraphics[width=\textwidth]{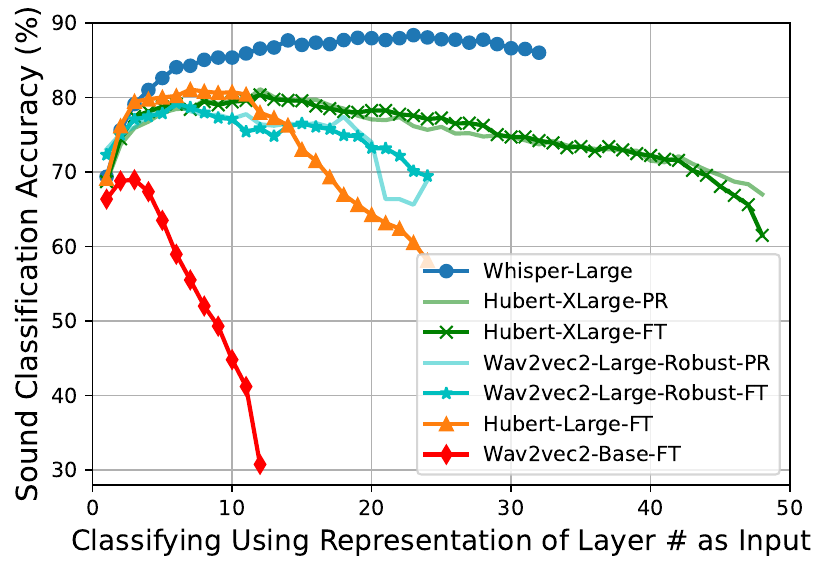}
\end{subfigure}
\squeezeup

\squeezeup

\caption{Surprisingly, the noise robustness of an ASR model correlates \textbf{positively} to the amount of general background sound (noise for ASR) information encoded in their intermediate representations. In the upper figure, we show Whisper is noticeably more robust (smaller word error rate increase) when speech (Librispeech) is contaminated with an increasing amount of background sounds from ESC-50~\cite{piczak2015esc}. In the lower figure, we show the intermediate representations of Whisper lead to the best linear probing sound classification accuracy on the same ESC-50 data, indicating Whisper encodes most background sound information. Unlike other models, Whisper encodes background sound information even in its deepest layer. PR=self-supervised pretrained; FT=PR and fine-tuned model.}
\label{fig:intro}
\squeezeup\squeezeup\squeezeup \squeezeup\squeezeup
\end{figure}

In this paper, we first show a counter-intuitive finding that while Whisper is robust against background sounds (noise for ASR), its audio representation is actually not noise-invariant, but instead encodes rich information of non-speech background sounds (shown in Figure~\ref{fig:intro} and discussed in detail in Section~\ref{sec:rep_ana}), indicating that the Whisper model does not learn a noise-invariant representation, but \emph{encodes} the noise type, and then recognize speech \emph{conditioned} on the noise type. 

One exciting application of the above finding is that we can build a \emph{unified} model for ASR and Audio Tagging (i.e., recognize general audio events) based on Whisper since it 1) is robust to noise, and 2) encodes rich general audio event information. Currently, ASR and audio tagging (AT) models are typically performed independently.  %, due a lack of model that meets these two conditions simultaneously. 
In many applications such as video transcribing, voice assistants, and hearing aid systems, we desire to get both spoken text and acoustic scene analysis from the audio, but running two systems is computationally expensive. In this work, we show that with $<$1\% extra computational cost, we can make Whisper recognizes audio events together with spoken text in a single forward pass. Our model achieves an mAP of 41.5 on AudioSet, which is slightly worse than standalone AT models, but is nevertheless over 40$\times$ faster.

\noindent\textbf{Related Work}: To the best of our knowledge, we are the first to report that a robust ASR actually learns a noise-variant representation; most previous work focuses on noise-invariant representations~\cite{van2009unsupervised,serdyuk2016invariant,sriram2018robust,liang2018learning,zhu2022noise}. For ASR and AT model unification, the closest works are~\cite{dinkel2022unikw,moritz2020all,narisetty2022joint,turian2022hear}. In~\cite{dinkel2022unikw}, a unified keyword spotting and audio tagging model is proposed, however, keyword spotting only considers up to 35 words and is a much simpler task than the large‐vocabulary continuous speech recognition task we are targeting. In \cite{moritz2020all,narisetty2022joint}, joint ASR and audio tagging/captioning training frameworks are proposed, but in this work, we show that Whisper already encodes rich general audio information even without any explicit audio tagging training. In~\cite{turian2022hear}, ASR representations are tested for the audio tagging task, but the overall performance is unsatisfactory.

\section{Whisper Robust ASR Model}

Whisper~\cite{radford2022robust} is a recently proposed robust ASR model that features a standard Transformer~\cite{vaswani2017attention}-based encoder-decoder architecture. The main novelty of Whisper is not its architecture, but its training data and training scheme. Specifically, the 680K-hour non-public training set contains audio-transcript pairs collected from the Internet with a very broad distribution of audio from many different environments, recording setups, speakers, and languages. Significant effort was made to filter out low-quality data. Compared with the most commonly used Librispeech (960 hours) and Libri-light (60K hours) data that are collected from audiobooks, the Whisper training data is much \emph{larger} and more \emph{diverse}, but also has noisy labels. We identify this as the main factor that differentiates Whisper from existing ASR models. During Whisper training, only text transcripts are used as supervision signals, no audio event labels are given. In this paper, we use the Whisper-Large model unless otherwise stated. Since Whisper is an encoder-decoder model, we only use the audio encoder part of Whisper for audio tagging, which consists of 32 Transformer layers with a dimension of 1280. 

%Due to space limitations, please refer to~\cite{radford2022robust} for more details about Whisper.

\section{Noise-Robust ASR Learns \\Noise-\emph{Variant} Representations}
\label{sec:rep_ana}

Thanks to the diverse 680K-hour training data, Whisper has been shown to be more robust under white and pub noise than its counterparts~\cite{radford2022robust}. We confirmed this point by evaluating Whisper and other state-of-the-art ASR models on Librispeech clean speech data that were contaminated with ESC-50~\cite{piczak2015esc} environmental sounds with various signal-to-noise ratios (SNRs). As shown in Figure~\ref{fig:intro} (upper), Whisper has superior performance. 

What is the noise-robust mechanism of Whisper? It is commonly believed that the representation of a robust ASR model should be noise-\emph{invariant}, and researchers often set noise-invariance as an explicit inductive bias for robust ASR (e.g., in~\cite{van2009unsupervised,serdyuk2016invariant,sriram2018robust,liang2018learning,zhu2022noise}). However, we, perhaps surprisingly, found that Whisper's representation is actually noise-\emph{variant} and encodes rich non-speech background sound information. 

Specifically, we froze the entire Whisper model and input audio samples from the ESC-50 environment sound dataset~\cite{piczak2015esc}. We then extracted the intermediate representation from every layer of Whisper and trained a linear layer on top of it to classify the sound class from 50 possible classes. If Whisper did not encode background sound information, or its representations were invariant to background sounds, the sound classification result would be low, and vice versa. As shown in Figure~\ref{fig:intro} (lower), the Whisper representations had the best ESC-50 sound classification accuracy compared to other SOTA ASR models, indicating that its representation encodes most background sound information. In addition, for all other ASR models, representations from deeper layers led to lower sound classification accuracies, showing that the models are learning to encode speech information, and ignore background sound information.  Whisper does not have this behavior, since representations from deeper layers also encode background sound information.

\begin{figure}[t]
\centering
\includegraphics[width=8.05cm]{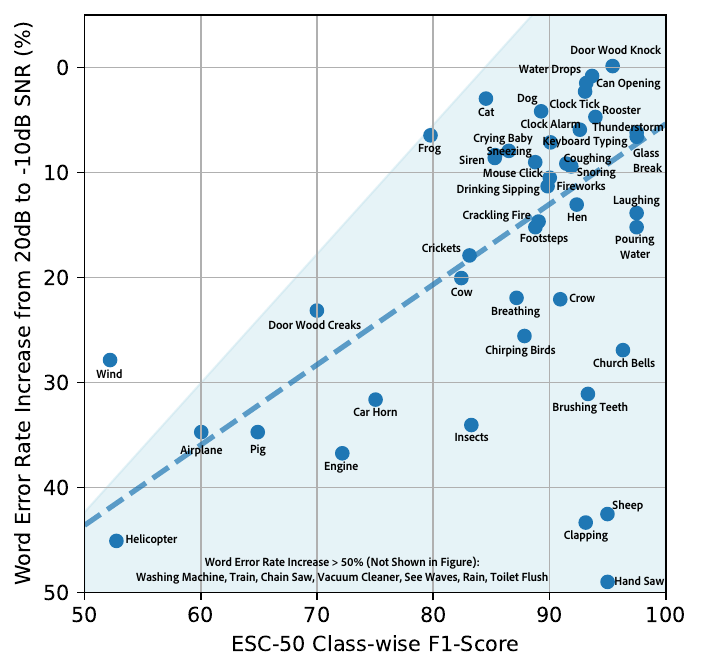}
\squeezeup\squeezeup
\caption{Class-wise analysis of the relationship between Whisper's robustness against a specific background sound class and its potential ability to recognize the sound. We measure Whisper robustness by its WER increase from clean speech (20dB SNR) to speech contaminated by the specific background sound from ESC-50 (-10dB SNR). The lower the WER increase, the more robust the model (Y-axis). We estimate the potential ability of Whisper to recognize the sound by training a linear layer on top of the Whisper encoder's last-layer representation for the sound classification task on the same ESC-50 dataset (without speech mixed-in, the Whisper model is frozen) and show the class-wise F1-score. The higher the F1-score, the better Whisper can potentially recognize the sound class (X-axis). Blue dashed line: we observe a positive correlation between Whisper's robustness against a background sound type and its potential ability to recognize it. Blue shading: we observe most sound classes lie in the right-bottom triangle area, indicating that Whisper is not robust to the type of sound if it cannot recognize the sound type. Right-bottom outliers: there are some background sounds that Whisper can potentially recognize but is not robust to, which is expected as some noises heavily overlap with the speech and are impossible to be robust to. In short, we find the potential ability to recognize a sound type is a necessary but not sufficient condition for Whisper to be robust to it.}
\squeezeup\squeezeup\squeezeup
\label{fig:cla_ana}
\end{figure}

The fact that Whisper is noise-robust while its representation encodes rich background sound information reveals that the robustness mechanism of Whisper is different from other ASR models (including wav2vec2-robust~\cite{hsu2021robust}).  Instead of learning a noise-invariant representation, it first \emph{encodes} the background sound and then transcribes text \emph{conditioned} on the type of noise. We confirmed this point by further checking the class-wise relationship between Whisper's robustness against a specific background sound class, and its potential ability to recognize the sound class in Figure~\ref{fig:cla_ana}.  We found there is indeed a positive correlation between them. Compared to \emph{noise-aware training}~\cite{seltzer2013investigation} that requires manually inputting noise type to the model, Whisper learns it directly from its massive 680K hour training set. 

Note that the discussion in this section is mostly based on Whisper, and our experiments do not indicate that noise-invariance does not help noise-robust ASR, nor that a noise-robust ASR's representation should be noise-variant. In fact, we believe encouraging noise-invariant representations~\cite{van2009unsupervised,serdyuk2016invariant,sriram2018robust,liang2018learning,zhu2022noise} is a practical solution in self-supervised learning or small data cases. Whisper training requires industry-level computational resources and is expensive. What we hope to convey is that a noise-robust ASR model does not have to learn a noise-invariant representation, and that there exist other ways to be noise-robust - a noise-conditioned model like Whisper can, and indeed does, work very well. 

\section{Unifying ASR and Audio Tagging Model}
\label{sec:whisper_at}

One exciting application of the finding in Section~\ref{sec:rep_ana} is that we are able to build a \emph{unified} model for ASR and Audio Tagging based on Whisper to recognize spoken text and background sounds (e.g., music, horn, etc) simultaneously, which is highly desirable in applications such as video transcribing, voice assistants, and hearing aid systems. Whisper is ideal as a backbone for such a unified model because 1) it is robust to background sounds, and 2) its intermediate representations encode rich general audio event information, which serves as a solid base for audio tagging. Nonetheless, the original Whisper does not output sound labels, so we need to train a model on top of Whisper intermediate representations to enable it to predict a sound class. Note that we intentionally do not modify the original weights of the Whisper model, but instead add new audio tagging layers on top of it so that the Whisper ASR ability is not changed and text and audio labels can be generated in a \emph{single} forward pass. We call this unified ASR and Audio Tagging model \emph{Whisper-AT}.

In previous sections, we applied a basic linear layer on the representation of a single layer for probing purposes. In this section, we discuss more advanced methods that lead to better audio tagging performance. 

\begin{figure}[t]
\centering
\includegraphics[width=6.0cm]{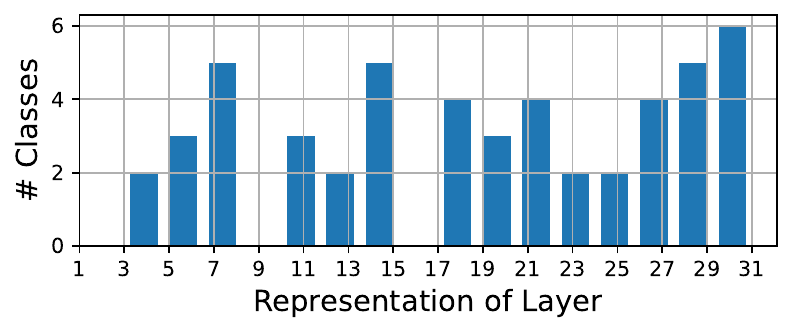}
\caption{Histrogram of the best Whisper representation layer (1-32) for the 50 ESC-50 sound classes. We train a linear layer on top of the representation of each of the 32 Whisper layers for ESC-50 sound classification, compute the class-wise F1-Score, and find the best representation layer for each sound class. Different sound classes get the best F1-score on representations of different layers.}
\label{fig:best_layer}
\squeezeup
\end{figure}

\begin{figure}[t]
\centering
\includegraphics[width=8.0cm]{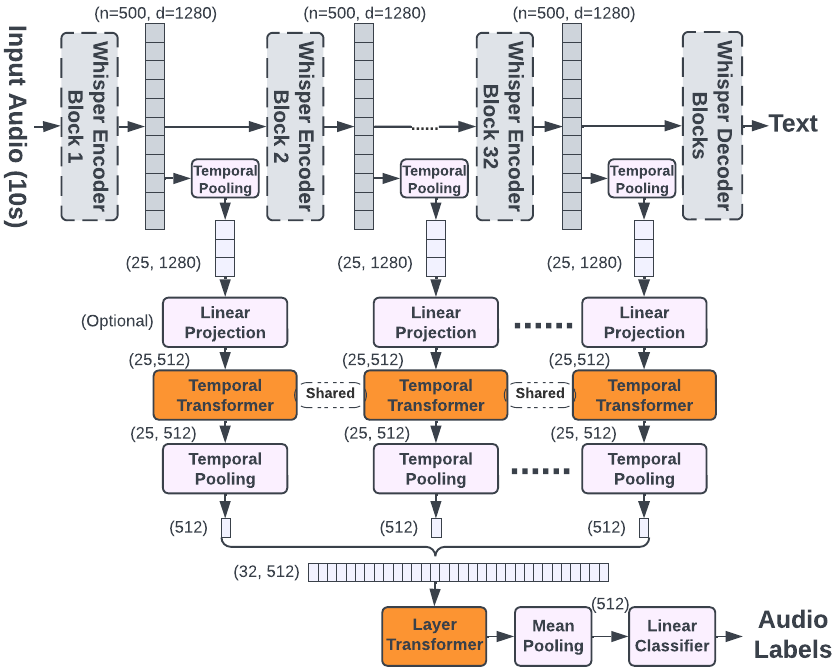}
\caption{The proposed time and layer-wise Transformer model.}
\label{fig:tltr}
\end{figure}

\begin{enumerate}
    \item \textbf{\texttt{Last-MLP}}: The most basic method, we first apply a temporal mean pooling over the last layer representation of Whisper and then apply a linear layer to map it to the prediction. 
    \item \textbf{\texttt{WA-MLP}}: As shown in Figure~\ref{fig:best_layer}, we find the last layer is not optimal for all sound classes. Thus we weighted average (WA) the representations from all layers and set the weight to be learnable before temporal mean pooling and linear layer, so this approach leverages representations from all layers. 
    \item \textbf{\texttt{WA-Tr}}: Temporal mean pooling removes all temporal details, and a single linear layer may be too simple for audio tagging. Therefore, we replace the linear layer of \texttt{WA-MLP} with a single-head temporal Transformer layer for this model.
    \item \textbf{\texttt{TL-Tr}}: Time and layer-wise Transformer (our main method, shown in Figure~\ref{fig:tltr}). Though weighted averaging leverage representation of all layers, all sound classes use a \emph{fixed} set of weights. In Figure~\ref{fig:best_layer}, we show that different sound classes achieve their best performance using different representation layers. Therefore, ideally, each class should have its own set of weights. This motivates us to build an attention mechanism over the \emph{layers}. Specifically, we apply another layer-wise Transformer to the output of the temporal Transformer.  
\end{enumerate}

\noindent\textbf{Efficient Design:} As the original goal of Whisper-AT is being more computationally efficient than two independent ASR and AT models, we aim to minimize the extra cost for audio tagging. Introducing a new Transformer layer in \texttt{WA-Tr} and \texttt{TL-Tr} is relatively expensive. Consider the complexity of Transformer is $O(d^2n+dn^2)$, where $d$ is the dimension and $n$ is the input length of the Transformer, for each 10-second input audio, the representations of each Whisper layer is in the shape of ($n$=500, $d$=1280). If the temporal and layer Transformer have the same $n$ and $d$ as Whisper, their computational cost is not negligible. Therefore, as illustrated in Figure~\ref{fig:tltr}, we propose the following efficient design: 1) We add a mean pooling layer to each Whisper representation to lower the time sequence length $n$ from 500 to 25; 2) We add an optional linear projection layer to lower $d$ from 1280 to 512 before audio tagging Transformers (denoted by \texttt{TL-Tr$_{512}$}); and 3) For \texttt{WA-Tr}, we first conduct weighted averaging and then apply a temporal Transformer, for \texttt{TL-Tr}, we use a single temporal Transformer for all layers. Thus both \texttt{WA-Tr} and \texttt{TL-Tr} only need one temporal Transformer. 

\section{Experiments}

As mentioned in Section~\ref{sec:whisper_at}, we intentionally freeze the weights of the original Whisper model, so the ASR performance of Whisper-AT is exactly the same as the original Whisper~\cite{radford2022robust}. Thus we only conduct experiments on the audio tagging task.

\begin{table*}[t]
\centering
\footnotesize
\caption{Audio tagging performance comparison on AS-20K, AS-2M (mAP), and ESC-50 (accuracy). $^\dagger$ASR backbone parameters and FLOPs are not included. $^\star$Speed-up = 1$/$FLOPs, compared with AST; FLOPs computed by \texttt{fvcore}~\cite{fvcore}. $^\ddagger$: labeled AS-2M data is also used. $^{**}$ AS-2M experiment is expensive, we skip it when AS-20K and ESC50 experiments already shown clear differences. End-to-End fine-tuning results are shown in grey text as the comparison is not exactly fair. }
\label{tab:main_res}
\begin{tabular}{lccccccc}
\hline
\multicolumn{1}{c}{Model}                & Training Setting                   & Method                            & AS-20K                      & AS-2M                        & ESC-50                       & AT \#Params$^\dagger$               & AT Speed-Up$^\dagger$$^\star$                               \\ \hline
\multicolumn{8}{l}{\textit{Existing Standalone Audio Tagging Models}}                                                                                                                                                                                                                    \\
{\color[HTML]{656565} AudioSet Baseline~\cite{gemmeke2017audio}} & {\color[HTML]{656565} Fine-Tuning} & {\color[HTML]{656565} \texttt{End-to-End}} & {\color[HTML]{656565} -}    & {\color[HTML]{656565} 31.4} & {\color[HTML]{656565} -}     & {\color[HTML]{656565} -}    & {\color[HTML]{656565} -}                    \\
{\color[HTML]{656565} AST~\cite{gong2021ast}}               & {\color[HTML]{656565} Fine-Tuning} & {\color[HTML]{656565} \texttt{End-to-End}} & {\color[HTML]{656565} 34.7} & {\color[HTML]{656565} 45.9} & {\color[HTML]{656565} 88.8}  & {\color[HTML]{656565} 87M}  & {\color[HTML]{656565} 1 \texttt{$\times$} (133G FLOPs)} \\
{\color[HTML]{656565} SSAST~\cite{gong2022ssast}}             & {\color[HTML]{656565} Fine-Tuning} & {\color[HTML]{656565} \texttt{End-to-End}} & {\color[HTML]{656565} 31.0} & {\color[HTML]{656565} -}    & {\color[HTML]{656565} 88.7}  & {\color[HTML]{656565} 87M}  & {\color[HTML]{656565} 1 \texttt{$\times$}}              \\
{\color[HTML]{656565} PANNs~\cite{kong2020panns}}             & {\color[HTML]{656565} Fine-Tuning} & {\color[HTML]{656565} \texttt{End-to-End}} & {\color[HTML]{656565} 27.8} & {\color[HTML]{656565} 43.9} & {\color[HTML]{656565} 94.7$^\ddagger$} & {\color[HTML]{656565} 81M}  & {\color[HTML]{656565} 2.5 \texttt{$\times$}}            \\
{\color[HTML]{656565} MAE-AST~\cite{baade2022mae}}           & {\color[HTML]{656565} Fine-Tuning} & {\color[HTML]{656565} \texttt{End-to-End}} & {\color[HTML]{656565} 30.6} & {\color[HTML]{656565} -}    & {\color[HTML]{656565} 90.0}  & {\color[HTML]{656565} 87M}  & {\color[HTML]{656565} 2.7 \texttt{$\times$}}            \\
{\color[HTML]{656565} Audio-MAE~\cite{huangmasked}}         & {\color[HTML]{656565} Fine-Tuning} & {\color[HTML]{656565} \texttt{End-to-End}} & {\color[HTML]{656565} 37.0} & {\color[HTML]{656565} 47.3} & {\color[HTML]{656565} 94.1}  & {\color[HTML]{656565} 87M}  & {\color[HTML]{656565} 2.7 \texttt{$\times$}}            \\ \hline
\multicolumn{8}{l}{\textit{Existing Automatic Speech Recognition Models}}                                                                                                                                                                                                                \\
Hubert X-Large~\cite{hsu2021hubert}                           & Frozen                             & \texttt{WA-MLP}                            & 18.5                        & - $^{**}$                           & 82.2                         & 0.7M                        & 195K \texttt{$\times$}                               \\
Hubert X-Large~\cite{hsu2021hubert}                           & Frozen                             & $\texttt{TL-Tr}_{\texttt{1280}}$                             & 20.2                        & -                           & 83.6                         & 40M                         & 5 \texttt{$\times$}                                   \\
wav2vec2-Large-Robust~\cite{hsu2021robust}                     & Frozen                             & \texttt{WA-MLP}                          & 18.1                        & -                           & 78.5                         & 0.5M                        & 244K \texttt{$\times$}                               \\
wav2vec2-Large-Robust~\cite{hsu2021robust}                     & Frozen                             & $\texttt{TL-Tr}_{\texttt{1024}}$                             & 20.2                        & -                           & 82.8                         & 26M                       & 17 \texttt{$\times$}                                  \\ \hline
\multicolumn{8}{l}{\textit{Whisper-AT}}                                                                                                                                                                                                                                                  \\
Whisper-Large                            & Frozen                             & \texttt{Last-MLP}                          & 20.6                        & 20.3                        & 87.0                         & 0.7M                        & 195K \texttt{$\times$}                               \\
Whisper-Large                            & Frozen                             & \texttt{WA-MLP}                            & 25.7                        & 32.4                        & 90.2                         & 0.7M                        & 195K \texttt{$\times$}                               \\
Whisper-Large                            & Frozen                             & \texttt{WA-Tr}                             & 32.1                        & 41.0                        & 91.0                         & 20M                         & 270 \texttt{$\times$}                                 \\
\rowcolor[HTML]{EFEFEF} 
Whisper-Large                            & Frozen                             & $\texttt{TL-Tr}_{\texttt{1280}}$                             & 33.0                        & 42.1                        & 91.1                         & 40M                         & 8 \texttt{$\times$}                                   \\
\rowcolor[HTML]{C0C0C0} 
Whisper-Large                            & Frozen                             & $\texttt{TL-Tr}_{512}$                        & 32.8                        & 41.5                        & 91.7                         & 7M                          & 42 \texttt{$\times$}                                  \\
{\color[HTML]{656565} Whisper-Large}     & {\color[HTML]{656565} Fine-Tuning} & {\color[HTML]{656565} \texttt{End-to-End}} & {\color[HTML]{656565} 34.7} & {\color[HTML]{656565} 45.7} & {\color[HTML]{656565} 90.0}  & {\color[HTML]{656565} 655M} & {\color[HTML]{656565} 0.4 \texttt{$\times$}}            \\
{\color[HTML]{656565} Whisper-Small}     & {\color[HTML]{656565} Fine-Tuning} & {\color[HTML]{656565} \texttt{End-to-End}} & {\color[HTML]{656565} 31.9} & {\color[HTML]{656565} 44.1} & {\color[HTML]{656565} 88.9}  & {\color[HTML]{656565} 94M}  & {\color[HTML]{656565} 2.5 \texttt{$\times$}}            \\ \hline
\end{tabular}
\squeezeup
\end{table*}

\subsection{Experiment Settings}

\textbf{Dataset:} We use AudioSet and ESC-50 datasets following standard evaluation protocols. AudioSet~\cite{gemmeke2017audio} is a collection of over 2 million 10-second audio clips excised from YouTube videos and labeled with the sounds that the clip contains from a set of 527 labels. We train our model with both the balanced training set (AS-20K) and full training set (AS-2M) and report mAP on the evaluation set. ESC-50~\cite{piczak2015esc} consists of 2,000 5-second environmental audio recordings organized into 50 classes; we evaluate our model using the official 5-fold cross-validation protocol.

\noindent\textbf{Hyper-Parameters:} We use the standard training pipeline in prior AT work~\cite{gong2021ast,gong2022ssast,gong_psla,nagrani2021attention}. For all experiments, we use a batch size of 48 and an Adam optimizer~\cite{adam}. For the proposed \texttt{TL-Tr$_{512}$} model, we use an initial learning rate of 2e-4, 1e-4, and 5e-4, and train the model for 30, 5, and 30 epochs for AS-20K, AS-2M, and ESC-50, respectively. For baseline methods, we search the learning rate to ensure a fair comparison. 

\subsection{Experiment Results}

We show the main results in Table~\ref{tab:main_res}. The key conclusions are: 

First, Whisper-AT is significantly stronger than Hubert X-Large~\cite{hsu2021hubert} and wav2vec2-Large-Robust~\cite{hsu2021robust} on audio tagging, demonstrating that Whisper is not only the most robust ASR model but also the strongest audio tagging backbone.

Second, comparing the four Whisper-AT models, the proposed \texttt{TL-Tr} model leads to the best performance with higher computational overhead. However, by projecting the Transformer dimension from 1280 to 512, $\texttt{TL-Tr}_{\texttt{512}}$ strikes a balance between performance and efficiency, as its FLOPs are less than 1\% of the Whisper ASR FLOPs yet it performs almost the same as $\texttt{TL-Tr}_{\texttt{1280}}$. In Table~\ref{tab:impact_d}, we further study the relationship between the audio tagging performance and Transformer dimension $d$ for \texttt{TL-Tr}. Even $\texttt{TL-Tr}_{\texttt{128}}$ provides reasonably good audio tagging performance, while its computational cost is almost free ($<$0.1\% FLOPs of the Whisper ASR FLOPs).

Third, Whisper-AT is slightly worse than SOTA standalone audio tagging models but is much more efficient. The proposed $\texttt{TL-Tr}_{512}$ achieves 32.8 mAP, 41.5 mAP, and 91.7 accuracy on AS-20K, AS-2M, and ESC-50, respectively, and is 42 times faster and 11 times smaller than AST~\cite{gong2021ast}.
Note that we target the cases that the user is already running an ASR and want to get additional audio labels, so we only compare the \emph{additional} cost for AT and do not include the cost of ASR in this comparison.

Fourth, how does Whisper perform in the end-to-end finetuning setting, and how does it compare to SOTA audio tagging models? We add a new Transformer layer on top of the Whisper encoder and train the entire model end-to-end (new layer uses a 10-100\texttt{$\times$} larger learning rate). For a fair comparison, we also test Whisper-Small which is of similar size to SOTA audio tagging models. We find Whisper-Small performs similarly with previous self-supervised pretrained models such as SSAST~\cite{gong2022ssast} and MAE-AST~\cite{baade2022mae} after fine-tuning.

%Please note that Whisper has an ``architecture disadvantage'' (e.g., Whisper uses 80 mel bins while AT models typically use 128) as its architecture is designed for ASR rather than AT.

\begin{figure}[t]
\centering
\includegraphics[width=8.0cm]{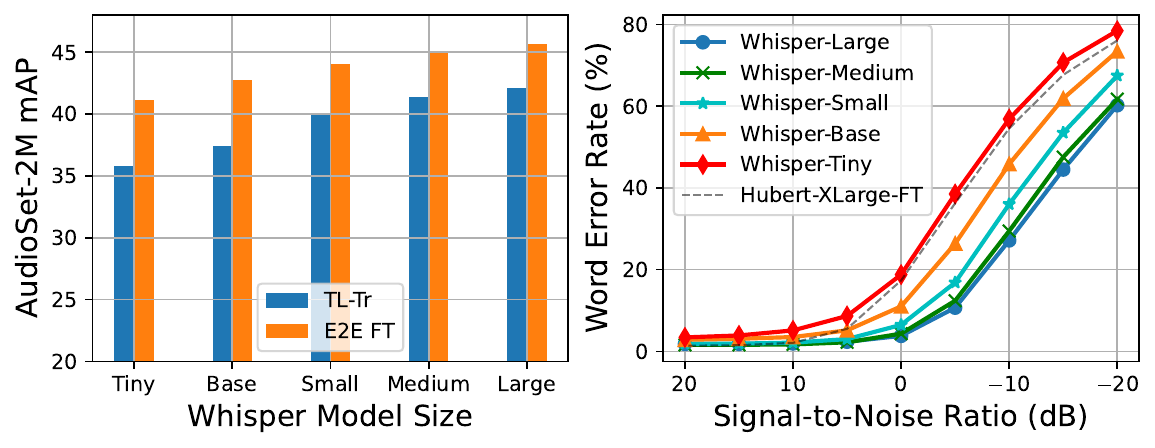}
\squeezeup\squeezeup\squeezeup
\caption{AS-2M audio tagging performance (left) and ASR robustness (right) of the Whisper model family.}
\label{fig:mdl_size}
\squeezeup\squeezeup
\end{figure}

Finally, we test the audio tagging performance of smaller Whisper models. As shown in Figure~\ref{fig:mdl_size}, smaller models have weaker audio tagging performance but the difference between Whisper-Small, Medium, and Large is minor. We also test the ASR noise-robustness of these models on speech contaminated by ESC50 background sounds; larger models are more robust. We again observe a positive correlation between ASR noise robustness and AT performance. In addition, Whisper-Base (74M parameters) is already more robust in ASR and stronger in audio tagging than Hubert-X-Large (964M parameters).

\begin{table}[t]
\centering
\footnotesize
\caption{Performance and efficiency impact of \texttt{TL-Tr} Transformer dimension $d$.}
\label{tab:impact_d}
\begin{tabular}{@{}ccccc@{}}
\toprule
Tr Dim $d$ & FLOPs (G) & \#Params (M) & AS-20K & ESC-50 \\ \midrule
128       & 0.31      & 0.6      & 30.0   & 91.4   \\
256       & 0.94      & 2.1      & 32.0   & 92.0   \\
512       & 3.17      & 7.2      & 32.8   & 91.7   \\
768       & 6.72      & 15.6     & 33.0   & 91.4   \\
1280      & 16.42     & 40.0     & 33.0   & 91.1   \\ \bottomrule
\end{tabular}
\end{table}

\section{Conclusion}

The Whisper ASR model revives the supervised learning scheme by using a massive and diverse training corpus. In this paper, we report an intriguing property of Whisper that while being very robust, the audio representation of Whisper is actually noise-variant and encodes rich background sound information. Based on this finding, we propose a unified audio tagging and ASR model called \emph{Whisper-AT}.  With only less than 1\% additional cost, Whisper-AT can recognize the background sound in addition to spoken text in a single forward pass.

\noindent\textbf{Acknowledgments:} This research is supported by the MIT-IBM Watson AI Lab.

\bibliographystyle{IEEEtran}
\bibliography{main}

\end{document}